\documentclass[12pt]{article}
\usepackage{myart}

\normalbaselines
\input tcilatex.tex

\begin{document}

\author{Yuri A. Rylov}
\title{Dynamical methods of investigations in application to the Schr\"{o}%
dinger particle}
\date{Institute for Problems in Mechanics, Russian Academy of Sciences \\
101-1 ,Vernadskii Ave., Moscow, 119526, Russia \\
email: rylov@ipmnet.ru\\
Web site: {$http://rsfq1.physics.sunysb.edu/\symbol{126}rylov/yrylov.htm$}\\
or mirror Web site: {$http://195.208.200.111/\symbol{126}rylov/yrylov.htm$}}
\maketitle

\begin{abstract}
Quantum systems are dynamic systems restricted by the principles of quantum
mechanics (linearity of dynamic equations, linear transformation of the wave
function etc.). One suggests to investigate the quantum systems simply as
dynamic systems, ignoring the quantum principles and constraints imposed by
them. Such dynamic methods of investigation appear to be more adequate and
effective, than the conventional quantum methods of investigation. Using
these methods, which ignore the quantum principles, one can overcome the
principal problem of quantum field theory: join of nonrelativistic quantum
principles with the relativity principles. Investigation of dynamic system $%
\mathcal{S}_{\mathrm{S}}$, described by the Schr\"{o}dinger equation, admits
one to prove that the Copenhagen interpretation is incompatible with quantum
mechanics formalism. Besides, it is shown that sometimes the application of
quantum principles leads to incorrect results.
\end{abstract}

\section{Introduction}

One uses new dynamical methods of investigation, which ignore the quantum
principles. Elimination of the quantum principles admits one to eliminate
the principal problem of QFT: \textit{join} of the \textit{relativity
principles} with the \textit{principles of quantum mechanics. }After this
elimination the relativistic quantum theory is constructed on the basis of
relativistic dynamics only, whereas the conventional approach uses in
addition the linearity of dynamic equations. This linearity is based on a
special gauge of the wave function. A use of such an artificial property of
the wave function for construction of the relativistic quantum theory seems
to be unjustified, although the linearity is very convenient, because the
linear equations are simple for solution. We show in the simple example of
Schr\"{o}dinger particle $\mathcal{S}_{\mathrm{S}}$ (dynamic system
described by the Schr\"{o}dinger equation), that the difference between the
quantum system and the classical one is purely dynamical, and the quantum
system may be described without a reference to quantum principles.

A fundamental physical theory $\mathcal{T}_{\mathrm{f}}$ must be a logical
structure. It means, that the theory contains a few fundamental
propositions. All predictions of the theory $\mathcal{T}_{\mathrm{f}}$ as
well as other propositions $\mathcal{P}_{\mathrm{i}}$ intermediate between
the fundamental propositions $\mathcal{P}_{\mathrm{f}}$ and experimental
data $\mathcal{D}_{\mathrm{e}}$ are deduced from the fundamental
propositions $\mathcal{P}_{\mathrm{f}}$ by means of logical reasonings and
mathematical calculations. Practical derivation of predictions and
explanations from the fundamental propositions $\mathcal{P}_{\mathrm{f}}$
may be difficult and complicated, because of long reasonings and complicated
calculations. In this case one uses a set of intermediate propositions $%
\mathcal{P}_{\mathrm{i}}$, which are valid in some region of physical
phenomena. Practical application of the intermediate propositions $\mathcal{P%
}_{\mathrm{i}}$ as some physical theory $\mathcal{T}_{\mathrm{c}}$ may
appear to be more effective and simpler, than a use of the fundamental
theory $\mathcal{T}_{\mathrm{f}}$.

In general, the intermediate propositions $\mathcal{P}_{\mathrm{i}}$ are to
be corollaries of the fundamental propositions $\mathcal{P}_{\mathrm{f}}$.
However, if it is impossible to discover the true fundamental propositions $%
\mathcal{P}_{\mathrm{f}}$, we may guess intermediate propositions $\mathcal{P%
}_{\mathrm{i}}$ and consider them as a curtailed physical theory $\mathcal{T}%
_{\mathrm{c}}$. The set of prescriptions $\mathcal{P}_{\mathrm{i}}$ is
chosen in such a way to explain some set of experimental data. The curtailed
theory $\mathcal{T}_{\mathrm{c}}$ is not a logical structure, it is simply a
list of prescriptions, which are not connected logically between themselves.
The logical structure appears, only if we discover and add the fundamental
propositions $\mathcal{P}_{\mathrm{f}}$, which generate this list of
prescriptions $\mathcal{P}_{\mathrm{i}}$ under some conditions. In this case
the intermediate propositions $\mathcal{P}_{\mathrm{i}}$ may be eliminated
from the formulation of the fundamental theory $\mathcal{T}_{\mathrm{f}}$,
because the propositions $\mathcal{P}_{\mathrm{i}}$ are corollaries of
fundamental propositions $\mathcal{P}_{\mathrm{f}}$.

Thus, we may use the curtailed physical theory $\mathcal{T}_{\mathrm{c}}$
instead of the fundamental physical theory $\mathcal{T}_{\mathrm{f}}$. For
instance, the axiomatic thermodynamics is a curtailed theory with respect to
kinetic theory, which may be considered to be a fundamental theory. The
conventional quantum mechanics is a curtailed theory, whereas the
corresponding fundamental theory is not known yet. Such an application of $%
\mathcal{T}_{\mathrm{c}}$ is possible only under those conditions (for
instance, in the nonrelativistic case), when the intermediate propositions $%
\mathcal{P}_{\mathrm{i}}$ are valid. However, extension of the curtailed
theory $\mathcal{T}_{\mathrm{c}}$ to other conditions (for instance, to the
relativistical case) may appear to be impossible, because the intermediate
propositions $\mathcal{P}_{\mathrm{i}}$ are only corollaries of the
fundamental propositions $\mathcal{P}_{\mathrm{f}}$ in the nonrelativistic
case. Under another conditions (for instance, in the relativistical case)
the fundamental propositions $\mathcal{P}_{\mathrm{f}}$ may generate another
intermediate propositions $\mathcal{P}_{\mathrm{i}}^{\ast }$, which does not
coincide with $\mathcal{P}_{\mathrm{i}}$ and cannot be obtained from $%
\mathcal{P}_{\mathrm{i}}$, because the set of $\mathcal{P}_{\mathrm{i}}$ is
simply a list of prescriptions (a list of corollaries of $\mathcal{P}_{%
\mathrm{f}}$), but not a logical structure. 

This fact is shown in the scheme of Figure.1. The direct way from the
conventional nonrelativistic quantum theory to the relativistic quantum
theory is very difficult, because the conventional nonrelativistic quantum
theory is a kind of curtailed theory. The quantum principles form the
essential part of this theory. In general, the quantum principles are to be
corollaries of fundamental propositions and of nonrelativistic constraints.
Unfortunately, we cannot be sure that the quantum principles are logical
corollaries of fundamental principles, because the fundamental propositions
were not known, when the quantum principles were formulated. The quantum
principles have been nicked, and we cannot be sure that they have nicked
correctly in all details. Even if the quantum principles has been nicked
correctly in all details, we cannot decide what is a corollary of
fundamental propositions and what is a corollary of nonrelativistic
constraints. To construct the relativistic quanum theory, we are to clean
out the nonrelativistic quantum theory from its nonrelativistic features.
The only reliable method of such a refinement is a return to the fundamental
propositions. Unfortunately, the fundamental propositions are not known. 

In particular, the particle production effect is the essential part of
relativistic quantum theory. However, this effect is present neither in
classical relativistic physics, nor in the nonrelativistic quantum
principles. The contemporary relativistic quantum theory takes into account
the particle production only formally, introducing the creation and
annihilation operators. At such a consideration of the particle production
effect the creation of particles is possible not only by pairs, but by
terns, by quaternaries, etc. It depends on the form of the Lagrangian. It
means that the particle production effect is taken into account on the
dynamical level, whereas it should be taken into account on the more
fundamental level, because in reality the particles are produced only by pairs. 
It is conditioned by the fact that the particle production is connected with the 
turn of the particle world line in time.

In this paper we try to obtain the fundamental proposition (the starting
point of the quantum theory) in the example of the Schr\"{o}dinger particle 

\section{Artificiality of conventional method of \newline
description}

The action for the Schr\"{o}dinger particle $\mathcal{S}_{\mathrm{S}}$ has
the form 
\begin{equation}
\mathcal{A}_{\mathrm{S}}\left[ \psi \right] =\int \left\{ \frac{i\hbar }{2}%
\left( \psi ^{\ast }\partial _{0}\psi -\partial _{0}\psi ^{\ast }\cdot \psi
\right) -\frac{\hbar ^{2}}{2m}\mathbf{\nabla }\psi ^{\ast }\mathbf{\nabla }%
\psi \right\} d^{4}x  \label{c2.1}
\end{equation}
where $\psi =\psi \left( t,\mathbf{x}\right) $ is a complex wave function.
The action carries out the complete description of the Schr\"{o}dinger
particle $\mathcal{S}_{\mathrm{S}}$, because it generates the dynamic
equation 
\begin{equation}
i\hbar \partial _{0}\psi =-\frac{\hbar ^{2}}{2m}\mathbf{\nabla }^{2}\psi ,
\label{c2.2}
\end{equation}
and corresponding canonical quantities: the 4-current $j^{k}$ and the
energy-momentum tensor $T_{l}^{k}$ 
\begin{equation}
j^{k}=\left\{ \rho ,\mathbf{j}\right\} =\left\{ \psi ^{\ast }\psi ,-\frac{%
i\hbar }{2m}\left( \psi ^{\ast }\mathbf{\nabla }\psi -\mathbf{\nabla }\psi
^{\ast }\cdot \psi \right) \right\}  \label{c2.3}
\end{equation}

Connection between the particle and the wave function is described by the
relations. 
\begin{equation}
\left\langle F\left( \mathbf{x},\mathbf{p}\right) \right\rangle =B\int \func{%
Re}\left\{ \psi ^{\ast }F\left( \mathbf{x},\mathbf{\hat{p}}\right) \psi
\right\} d\mathbf{x,\qquad \hat{p}}=-i\hbar \mathbf{\mathbf{\nabla },\qquad }%
B=\left( \int \psi ^{\ast }\psi d\mathbf{x}\right) ^{-1}  \label{c2.4}
\end{equation}
which define the mean value $\left\langle F\left( \mathbf{x},\mathbf{p}%
\right) \right\rangle $ of any function $F\left( \mathbf{x},\mathbf{p}%
\right) $ of position $\mathbf{x}$ and momentum $\mathbf{p}$. We shall refer
to these relations together with the restrictions imposed on its
applications as the quantum principles, because von Neumann \cite{N32} has
shown, that all proposition of quantum mechanics can be deduced from
relations of this type.

Setting $\hbar =0$ in the action (\ref{c2.1}), we hope to obtain a
description of classical particle. Instead, the description disappears, and
in this sense the description (\ref{c2.1}) is artificial.

\section{Natural description of the Schr\"{o}dinger particle}

To obtain the natural description of the Schr\"{o}dinger\textbf{\ }particle,
we produce the change of variables (another gauge of the wave function
phase) 
\begin{equation}
\psi \rightarrow \Psi _{b}:\quad \psi =\left\vert \Psi _{b}\right\vert \exp
\left( \frac{b}{\hbar }\log \frac{\Psi _{b}}{\left\vert \Psi _{b}\right\vert 
}\right) \quad b=\text{const}\neq 0  \label{c3.1}
\end{equation}
in the action (\ref{c2.1}). We obtain 
\[
\mathcal{A}_{\mathrm{S}}\left[ \Psi _{b}\right] =\int \left\{ \frac{ib}{2}%
\left( \Psi _{b}^{\ast }\partial _{0}\Psi _{b}-\partial _{0}\Psi _{b}^{\ast
}\cdot \Psi _{b}\right) -\frac{b^{2}}{2m}\mathbf{\nabla }\Psi _{b}^{\ast }%
\mathbf{\nabla }\Psi _{b}\right. 
\]
\begin{equation}
+\left. \frac{b^{2}}{2m}\left( \mathbf{\nabla }\left\vert \Psi
_{b}\right\vert \right) ^{2}-\frac{\hbar ^{2}}{2m}\left( \mathbf{\nabla }%
\left\vert \Psi _{b}\right\vert \right) ^{2}\right\} dtd\mathbf{x}
\label{c3.2}
\end{equation}
This change of variables leads to the replacement $\hbar \rightarrow b$ and
to appearance of two nonlinear terms which compensate each other, if $%
b=\hbar $.

The dynamic equation becomes to be nonlinear, if $b^{2}\neq \hbar ^{2}$ 
\begin{equation}
ib\partial _{0}\Psi _{b}=-\frac{b^{2}}{2m}\mathbf{\nabla }^{2}\Psi _{b}-%
\frac{\hbar ^{2}-b^{2}}{8m}\left( \frac{\left( \mathbf{\nabla }\rho \right)
^{2}}{\rho ^{2}}+2\mathbf{\nabla }\frac{\mathbf{\nabla }\rho }{\rho }\right)
\Psi _{b},  \label{c3.4}
\end{equation}
\begin{equation}
\rho =\Psi _{b}^{\ast }\Psi _{b},\qquad \mathbf{j}=-\frac{ib}{2m}\left( \Psi
_{b}^{\ast }\mathbf{\nabla }\Psi _{b}-\mathbf{\nabla }\Psi _{b}^{\ast }\cdot
\Psi _{b}\right)  \label{c3.5}
\end{equation}
However, the description becomes to be natural in the sense, that after
setting $\hbar =0$, the action $\mathcal{A}_{\mathrm{S}}\left[ \Psi _{b}%
\right] $ turns into the action 
\begin{equation}
\mathcal{A}_{\mathrm{Scl}}\left[ \Psi _{b}\right] =\int \left\{ \frac{ib}{2}%
\left( \Psi _{b}^{\ast }\partial _{0}\Psi _{b}-\partial _{0}\Psi _{b}^{\ast
}\cdot \Psi _{b}\right) -\frac{b^{2}}{2m}\mathbf{\nabla }\Psi _{b}^{\ast }%
\mathbf{\nabla }\Psi +\frac{b^{2}}{2m}\left( \mathbf{\nabla }\left\vert \Psi
_{b}\right\vert \right) ^{2}\right\} dtd\mathbf{x}  \label{c3.6a}
\end{equation}
which describes the statistical ensemble $\mathcal{E}\left[ \mathcal{S}_{%
\mathrm{d}}\right] $ of free classical particles $\mathcal{S}_{\mathrm{d}}$.
The action $\mathcal{A}_{\mathcal{E}\left[ \mathcal{S}_{\mathrm{d}}\right] }$
for this statistical ensemble can be represented in the form 
\begin{equation}
\mathcal{A}_{\mathcal{E}\left[ \mathcal{S}_{\mathrm{d}}\right] }\left[ 
\mathbf{x}\right] =\int \frac{m}{2}\left( \frac{d\mathbf{x}}{dt}\right)
^{2}dtd\mathbf{\xi }  \label{c3.7}
\end{equation}
where $\mathbf{x}=\mathbf{x}\left( t,\mathbf{\xi }\right) $ is a 3-vector
function of independent variables $t,\mathbf{\xi =}\left\{ \xi _{1,}\xi
_{2},\xi _{3}\right\} $. The variables (Lagrangian coordinates) $\mathbf{\xi 
}$ label particles $\mathcal{S}_{\mathrm{d}}$ of the statistical ensemble $%
\mathcal{E}\left[ \mathcal{S}_{\mathrm{d}}\right] $. The statistical
ensemble $\mathcal{E}\left[ \mathcal{S}_{\mathrm{d}}\right] $ is a dynamical
system of the hydrodynamical type. One can show that the dynamic system,
described by the action $\mathcal{A}_{\mathrm{Scl}}\left[ \Psi _{b}\right] $
(\ref{c3.6a}) is a partial case (irrotational flow) of the dynamic system $%
\mathcal{E}\left[ \mathcal{S}_{\mathrm{d}}\right] $ \cite{R99}.

Connection between the Schr\"{o}dinger equation and hydrodynamical
description is well known \cite{M26,B52}. But a connection between the
description in terms of wave function and the hydrodynamic description was
one-way. One can transit from the Schr\"{o}dinger equation to the
hydrodynamic equations, but one cannot transit from hydrodynamic equations
to the description in terms of the wave function, because one \textit{needs}
to \textit{integrate} hydrodynamic equations. Indeed, the Schr\"{o}dinger
equation consists of two real first order equations for the density $\rho $
and the phase $\varphi $, whereas the system of the hydrodynamical equations
consists of four first order equations for the density $\rho $ and for the
velocity $\mathbf{v}$. To obtain four hydrodynamic equations one needs to
take gradient of the equation for the phase $\varphi $. On the contrary, if
we transit from the hydrodynamic description to the description in terms of
the wave function, we are to integrate hydrodynamic equations. In the
general case this integration was not known for a long time.

Change of variables, leading from the action $\mathcal{A}_{\mathcal{E}\left[ 
\mathcal{S}_{\mathrm{d}}\right] }\left[ \mathbf{x}\right] $ to the action $%
\mathcal{A}_{\mathrm{Scl}}\left[ \Psi _{b}\right] $ contains integration
(see \cite{R99} or mathematical appendices to papers \cite{R2004,R2004a}).
The constant $b$ in the action $\mathcal{A}_{\mathrm{Scl}}\left[ \Psi _{b}%
\right] $ is an arbitrary constant of integration (gauge constant).
Arbitrary integration functions are "hidden" inside the wave function $\Psi
_{b}$. Thus, the limit of Schr\"{o}dinger particle (\ref{c3.2}) at $\hbar
\rightarrow 0$ is a statistical ensemble $\mathcal{E}\left[ \mathcal{S}_{%
\mathrm{d}}\right] $, but not an individual particle $\mathcal{S}_{\mathrm{d}%
}$. It means, that the \textit{wave\ function\ describes\ a statistical\
ensemble of\ particles,\ but\ not\ an\ individual particle}, and Copenhagen\
interpretation, where the wave function describes an individual particle,
is\ \textit{incompatible}\ with the\ quantum\ mechanics\ formalism.

A use of Copenhagen\ interpretation does not generate any problems, until we
consider mathematical formalism of QM, because in the framework of this
formalism we have the only object of investigation. It is of no importance
what is the name of the investigated object. But at the consideration of the
measurement we have two different kinds of measurement:

\begin{enumerate}
\item Individual measurement ($S$-measurement), which is produced over the
individual stochastic particle $\mathcal{S}_{\mathrm{st}}$.

\item Massive measurement ($M$-measurement), which is produced over the
statistical ensemble $\mathcal{E}\left[ \mathcal{S}_{\mathrm{st}}\right] $,
or over the statistical average particle $\left\langle \mathcal{S}%
\right\rangle $.
\end{enumerate}

\section{Statistical ensemble as a starting point of \newline
quantum theory}

One can show, that the Schr\"{o}dinger particle $\mathcal{S}_{\mathrm{S}}$
is a special case of the statistical ensemble $\mathcal{E}\left[ \mathcal{S}%
_{\mathrm{st}}\right] $ of stochastically moving free particles $\mathcal{S}%
_{\mathrm{st}}$, i.e. the action (\ref{c3.2}) for $\mathcal{S}_{\mathrm{S}}$
can be obtained from the action for the statistical ensemble $\mathcal{E}%
\left[ \mathcal{S}_{\mathrm{st}}\right] $ by means of a proper change of
variables \cite{R99}.

The statistical ensemble $\mathcal{E}\left[ \mathcal{S}_{\mathrm{st}}\right] 
$ of \textit{stochastic} particles $\mathcal{S}_{\mathrm{st}}$ is a dynamic
system, described by the action 
\begin{equation}
\mathcal{A}_{\mathcal{E}\left[ \mathcal{S}_{\mathrm{st}}\right] }\left[ 
\mathbf{x,u}_{\mathrm{df}}\right] =\int \left\{ \frac{m}{2}\left( \frac{d%
\mathbf{x}}{dt}\right) ^{2}+\frac{m}{2}\mathbf{u}_{\mathrm{df}}^{2}-\frac{%
\hbar }{2}\mathbf{\nabla u}_{\mathrm{df}}\right\} dtd\mathbf{\xi }
\label{c4.1}
\end{equation}
where $\mathbf{u}_{\mathrm{df}}=\mathbf{u}_{\mathrm{df}}\left( t,\mathbf{x}%
\right) $ is the diffusion velocity, describing the mean value of the
stochastic component of the velocity, whereas $\frac{d\mathbf{x}}{dt}\left(
t,\mathbf{\xi }\right) $ describes the regular component of the particle
velocity, and $\mathbf{x}=\mathbf{x}\left( t,\mathbf{\xi }\right) $ is a
3-vector function of independent variables $t,\mathbf{\xi =}\left\{ \xi
_{1,}\xi _{2},\xi _{3}\right\} $. The variables $\mathbf{\xi }$ label
stochastic particles $\mathcal{S}_{\mathrm{st}}$, constituting the
statistical ensemble. The operator $\mathbf{\nabla }$ is defined in the
space of coordinates $\mathbf{x}$.\textbf{\ }Dynamic equations have the form

\begin{equation}
\frac{\delta \mathcal{A}_{\mathcal{E}\left[ \mathcal{S}_{\mathrm{st}}\right]
}}{\delta \mathbf{x}}=-m\frac{d^{2}\mathbf{x}}{dt^{2}}+\mathbf{\nabla }%
\left( \frac{m}{2}\mathbf{u}_{\mathrm{df}}^{2}-\frac{\hbar }{2}\mathbf{%
\nabla u}_{\mathrm{df}}\right) =0  \label{c4.2}
\end{equation}
\begin{equation}
\frac{\delta \mathcal{A}_{\mathcal{E}\left[ \mathcal{S}_{\mathrm{st}}\right]
}}{\delta \mathbf{u}_{\mathrm{df}}}=m\rho \mathbf{u}_{\mathrm{df}}+\frac{%
\hbar }{2}\mathbf{\nabla }\rho =0,  \label{c4.3}
\end{equation}
\begin{equation}
\rho =\left[ \frac{\partial \left( x^{1},x^{2},x^{3}\right) }{\partial
\left( \xi _{1},\xi _{2},\xi _{3}\right) }\right] ^{-1}=\frac{\partial
\left( \xi _{1},\xi _{2},\xi _{3}\right) }{\partial \left(
x^{1},x^{2},x^{3}\right) }  \label{c4.4}
\end{equation}
Resolving (\ref{c4.3}) with respect to $\mathbf{u}_{\mathrm{df}}$ in the
form 
\begin{equation}
\mathbf{u}_{\mathrm{df}}=-\frac{\hbar }{2m}\mathbf{\nabla }\ln \rho ,
\label{c4.5}
\end{equation}
and eliminating $\mathbf{u}_{\mathrm{df}}$ from equation (\ref{c4.2}), we
obtain the dynamic equations of the hydrodynamical type 
\begin{equation}
m\frac{d^{2}\mathbf{x}}{dt^{2}}=-\mathbf{\nabla }U\left( \rho ,\mathbf{%
\nabla }\rho \right) ,\qquad U\left( \rho ,\mathbf{\nabla }\rho \right) =%
\frac{\hbar ^{2}}{8m}\left( \frac{\left( \mathbf{\nabla }\rho \right) ^{2}}{%
\rho ^{2}}-2\frac{\mathbf{\nabla }^{2}\rho }{\rho }\right)  \label{c4.7}
\end{equation}

Hydrodynamic equations (\ref{c4.7}) may be written in terms of the wave
function \cite{R99,R2004,R2004a}. The proper change of the variables
together with \textit{integration} turns the action (\ref{c4.1}) for $%
\mathcal{E}\left[ \mathcal{S}_{\mathrm{st}}\right] $ into the action
containing the quantum constant $\hbar $ and an arbitrary integration
constant $b$.

The wave function is not a specific quantum quantity. The wave function is a
method of description of \textit{any ideal fluid}. Quantum and classical
dynamic systems distinguish \textit{dynamically} (by the form of their
actions). The form of description (in terms of wave function, or in terms of
position and momentum of the particle) is of no importance.

Considering the statistical ensemble of stochastic particles as a starting
point of the quantum mechanics, \textit{one does not need quantum principles}%
, because the interpretation is carried out directly via the particles of
the statistical ensemble. Besides, the quantum theory turns into a
consistent statistical theory, where there are two sorts of particles: (1)
individual stochastic particle $\mathcal{S}_{\mathrm{st}}$ and (2)
statistical average particle $\left\langle \mathcal{S}_{\mathrm{st}%
}\right\rangle $, which is the statistical ensemble, normalized to one
particle. Formalism of quantum mechanics deals only with the statistical
average particle $\left\langle \mathcal{S}_{\mathrm{st}}\right\rangle $. All
predictions of QM relate only to $\left\langle \mathcal{S}_{\mathrm{st}%
}\right\rangle $. In accordance with two sorts of particles we have two
sorts of measurements: (1) $S$-measurement produced over $\mathcal{S}_{%
\mathrm{st}}$ and (2) $M$-measurement produced over $\left\langle \mathcal{S}%
_{\mathrm{st}}\right\rangle $. The two kinds of measurements have different
properties and one may not confuse them (see for details \cite{R2002}).

The Copenhagen interpretation meets the difficulties, when it tries to test
predictions of mathematical formalism in single experiments. For instance,
there exists the problem of the mechanism of the wave function reduction at
a single experiment. Another problem concerns the two-slit experiment. How
can an individual particle pass through two slits at once? The physical
journals publish discussions concerning problems of quantum measurements.
For instance, such a discussion was declared in 2002 by the journal Uspekhi
Fizicheskich Nauk. These problems cannot be solved in the framework of the
Copenhagen interpretation, which does not distinguish between the individual
particle $\mathcal{S}$ and the statistically average particle $\left\langle 
\mathcal{S}\right\rangle $. Confusion of two different objects, having
different properties generates difficulties and paradoxes. The wave function
does not describe the state of individual particle $\mathcal{S}$, and it is
meaningless to ask, how the wave function changes at a single measurement ($%
S $-measurement). At the massive experiment ($M$-measurement) we obtain a
distribution $F\left( R^{\prime }\right) $ of the measured quantity $%
\mathcal{R}$, but not a single value $R^{\prime }$ of the measured quantity.
At such a situation it is useless to ask, how the obtained result $R^{\prime
}$ influences on the the state of the statistical ensemble (statistical
average particle). Finally, we may define the third type of measurement ($SM$%
-measurement): the massive measurement of the quantity $\mathcal{R}$ leading
to a definite value $R^{\prime }$ of the measured quantity $\mathcal{R}$.
The $SM$-measurement is the $M$-measurement leading to a distribution $%
F\left( R^{\prime }\right) $, accompanied by a selection of those particles,
where result of $S$-measurement is $R^{\prime }$. Uniting all particles with
the measured value $R^{\prime }$ in one statistical ensemble $\mathcal{E}%
_{R^{\prime }}$, we can put the question about the wave function of $%
\mathcal{E}_{R^{\prime }}$. Of course, the wave function $\psi _{R^{\prime
}} $ of $\mathcal{E}_{R^{\prime }}$ does not coincide, in general, with the
initial wave function $\psi $, and this change of the wave function is
considered as a reduction of the wave function. The origin of the reduction
is quite transparent. It is the selection, which is produced to obtain the
same value $R^{\prime }$ of the measured quantity for all particles of the
statistical ensemble. Thus, the problems of reduction are conditioned by the
confusion of concepts of the individual particle $\mathcal{S}$ and the
statistical average particle $\left\langle \mathcal{S}\right\rangle $, which
takes place at the Copenhagen interpretation (see for details \cite{R2002}).

As concerns the particle, passing through two slits simultaneously, it is a
reasonable property of the statistical average object. It is a pure
statistical property, which has nothing to do with quantum properties.
Individual particle $\mathcal{S}$ can pass either through one slit, or
through another, whereas the statistical average particle $\left\langle 
\mathcal{S}\right\rangle $ can pass through both slits simultaneously.
(Compare, individual person is either a man, or a woman, whereas the
statistical average person is a hermaphrodite (half-man half-woman), and
there are no quantum mechanical properties here).

Using statistical ensemble as a starting point, we can test validity the
quantum principles. The quantum principles or their conventional
interpretation appear to be wrong in some cases. For instance, the momentum
distribution $w\left( \mathbf{p}\right) =\left\vert \left\langle \mathbf{p}%
|\psi \right\rangle \right\vert ^{2}$ is in reality a distribution over mean
momenta $\left\langle \mathbf{p}\right\rangle $. Let us illustrate the
difference between the two distribution in the example of ideal gas, whose
state may be described by the wave function. In the case, when the gas moves
with the constant velocity $\mathbf{u}$, the wave function has the form 
\begin{equation}
\psi \left( \mathbf{x}\right) =A_{1}e^{\frac{i}{\hbar }m\mathbf{ux}},\qquad
A_{1}=\text{const}  \label{h2.1}
\end{equation}
where $m$ is the mass of a molecule. Corresponding momentum distribution has
the form 
\begin{equation}
w\left( \mathbf{p}\right) =\left\vert \left\langle \mathbf{p}|\psi
\right\rangle \right\vert ^{2}=A\delta \left( \mathbf{p}-m\mathbf{u}\right)
,\qquad A=\text{const}  \label{h2.2}
\end{equation}
This distribution does not coincides with the Maxwell momentum distribution 
\begin{equation}
F\left( \mathbf{x},\mathbf{p}\right) d\mathbf{p}=\frac{1}{\left( 2\pi
mkT\right) ^{3/2}}\exp \left\{ -\frac{\left( \mathbf{p}-m\mathbf{u}\right)
^{2}}{2mkT}\right\} d\mathbf{p}  \label{h2.3}
\end{equation}
which depends on the gas temperature $kT$.

Let us divide the space into small cells $V_{i}$ and calculate the mean
momentum $\left\langle \mathbf{p}\right\rangle _{i}$ of molecule in each
cell $V_{i}$. In the given case we obtain in the $i$th cell $\left\langle 
\mathbf{p}\right\rangle _{i}=m\mathbf{u}$. All mean momenta $\left\langle 
\mathbf{p}\right\rangle _{i}$ form the mean momenta distribution (\ref{h2.2}%
). The mean momenta $\left\langle \mathbf{p}\right\rangle _{i}$ correlate
with the position $\mathbf{x}_{i}$ of the cell $V_{i}$, and mutual
distribution over position $\mathbf{x}$ and momentum $\left\langle \mathbf{p}%
\right\rangle $ appears to be impossible. More detail consideration one can
find in \cite{R2004b}.

Description of "classical particle " $\mathcal{S}_{\mathrm{Scl}}=\mathcal{E}%
\left[ \mathcal{S}_{\mathrm{d}}\right] $ in the "quantum language" is
realized by the action. 
\begin{equation}
\mathcal{A}_{\mathrm{Scl}}\left[ \psi ,\psi ^{\ast }\right] =\int \left\{ 
\frac{i\hbar }{2}\left( \psi ^{\ast }\partial _{0}\psi -\partial _{0}\psi
^{\ast }\cdot \psi \right) -\frac{\hbar ^{2}}{2m}\mathbf{\nabla }\psi ^{\ast
}\mathbf{\nabla }\psi +\frac{\hbar ^{2}}{2m}\left( \mathbf{\nabla }%
\left\vert \psi \right\vert \right) ^{2}\right\} dtd\mathbf{x}  \label{c4.8}
\end{equation}
where we use the quantum constant $\hbar $ instead of arbitrary constant $b$%
. 
\begin{equation}
i\hbar \partial _{0}\psi +\frac{\hbar ^{2}}{2m}\mathbf{\nabla }^{2}\psi -%
\frac{\hbar ^{2}}{8m}\left( 2\frac{\mathbf{\nabla }^{2}\rho }{\rho }-\frac{%
\left( \mathbf{\nabla }\rho \right) ^{2}}{\rho ^{2}}\right) \psi =0\qquad
\rho =\psi ^{\ast }\psi  \label{c4.9}
\end{equation}

Description of "quantum particle" $\mathcal{S}_{\mathrm{S}}=\mathcal{E}\left[
\mathcal{S}_{\mathrm{st}}\right] $ in the "classical language" is realized
by equations

\begin{equation}
\frac{d\mathbf{p}}{dt}=-\mathbf{\nabla }U\left( \rho ,\mathbf{\nabla }\rho
\right) ,\qquad \frac{d\mathbf{x}}{dt}=\frac{\mathbf{p}}{m},  \label{c4.11}
\end{equation}
where $U$ is defined by the second relation (\ref{c4.7}).

Dynamic equations (\ref{c4.11}) are the \textit{partial} differential
equations, because $\rho $ contains derivatives with respect to $\xi
_{\alpha }$, $\alpha =1,2,3.$

Description of "classical particle" $\mathcal{S}_{\mathrm{Scl}}=\mathcal{E}%
\left[ \mathcal{S}_{\mathrm{d}}\right] $ in the "classical language" has the
form 
\begin{equation}
\frac{d\mathbf{p}}{dt}=0,\qquad \frac{d\mathbf{x}}{dt}=\frac{\mathbf{p}}{m}
\label{c4.12}
\end{equation}
where $\mathbf{x}=\mathbf{x}\left( t,\mathbf{\xi }\right) $, $\mathbf{p}=%
\mathbf{p}\left( t,\mathbf{\xi }\right) $. Dynamic equations (\ref{c4.12})
are \textit{ordinary} differential equations.

Describing quantum system $\mathcal{S}$ and corresponding classical system $%
\mathcal{S}_{\mathrm{cl}}$ in the classical language, we recognize that
dynamic equations for the quantum system $\mathcal{S}$ are \textit{partial}
differential equations, whereas dynamic equations for its classical
approximation $\mathcal{S}_{\mathrm{cl}}$ are \textit{ordinary} differential
equations.

The difference between the ordinary and partial differential equations is
dynamical (but not quantum). This invariant difference is a foundation of
the procedure of the dynamical disquantization (transition to classical
approximation).

Dynamic\ \ disquantization is the procedure of projecting of derivatives
onto the direction of 4-current $j^{k}$%
\[
\partial ^{l}\rightarrow \partial _{||}^{l}=\frac{j^{l}j^{k}}{j^{s}j_{s}}%
\partial _{k},\qquad l=0,1,2,3, 
\]
To\ obtain\ ordinary\ differential\ equations, one$\ $needs\ to\ project\
derivatives\ onto\ one\ direction.

The dynamic disquantization admits one to \textit{formalize} the transition
to the classical description. Being applied to Schr\"{o}dinger particle $%
\mathcal{S}_{\mathrm{S}}$, the dynamic disquantization transform $\mathcal{S}%
_{\mathrm{S}}=\mathcal{E}\left[ \mathcal{S}_{\mathrm{st}}\right] $ into the
statistical ensemble $\mathcal{E}\left[ \mathcal{S}_{\mathrm{d}}\right] $ of
free classical particles $\mathcal{S}_{\mathrm{d}}$.

The procedure of dynamic disquantization is determined by the dynamic system 
\textit{completely} and does not refer to the quantum constant and the
quantum principles.

\section{Concluding remarks}

We have analysed application of dynamical methods to the Schr\"{o}dinger
particle and have obtained results, which cannot be obtained by means of
conventional axiomatic methods. Analogous results are obtained at
application of dynamical methods to other quantum systems. Application of
dynamical methods to investigation of quantum systems admits one to
eliminate quantum principles. The number of fundamental propositions is
reduced essentially. The quantum theory ceases to be a list of prescriptions
and becomes a logical structure.

In the conventional quantum theory the allness of the quantum constant $%
\hbar $ is explained by the quantum nature of all physical phenomena. In
particular, it means that all physical fields should be quantized.
Application of dynamical methods and elimination of quantum principles
suppose some primordial stochastic motion of free particles with dependence
of the stochasticity intensity on the particle mass. The universal character
of the quantum constant and the particle motion stochasticity may be
explained freely by the space-time properties, when the quantum constant is
a parameter of the space-time geometry \cite{R1991,R2005}. In this case not
all physical fields have the quantum nature. Dynamic equations for the
metric fields (gravitational and electromagnetic) do not contain the quantum
constant $\hbar $. There is no necessity to quantize these fields.

It is well known that the electromagnetic field is absorbed and emitted in
the form of quanta of energy $\hbar \omega $. However, this fact does not
mean that the electromagnetic field exists in the form of quanta. There are
no experiments, which show directly or indirectly that the electromagnetic
field exists in the form of quanta. (At least, such experiments are unknown
for us). Absorbtion and emission of the electromagnetic field in the form of
quanta may be easily explained by the quantum properties of the absorbtion
devices and the emission ones. The same relates to the gravitational field.
All attempts of its quantization appear to be unsuccessful.

Quantum and classical dynamic systems distinguish \textit{only} dynamically,
i.e. by their dynamic equations, but not by their enigmatic quantum
properties. Linearity of the conventional axiomatic methods is simple and
attractive, but these methods are founded on nonrelativistic quantum
principles, and we cannot be sure that these methods may be applied in the
relativistic case.

There is a hope that application of dynamical methods in the relativistic
case will be successful. First, extension to the relativistic case is
produced on the logical basis, but not by means of incident hypotheses.
Second, in the relativistic case the diffusion velocity\ $\mathbf{u}_{%
\mathrm{df}}$ turns to the relativistic field, responsible for pair
production, whereas neither quantum principles nor classical relativistic
dynamics can describe effect of pair production, which is the principle
effect of the high energy physics.

\end{document}